\documentclass[journal]{IEEEtran}
\usepackage[T1]{fontenc} 
\usepackage{url}
\usepackage{amsmath}
\usepackage{graphicx}
\usepackage[justification=centering]{caption}
\usepackage{cite}
\usepackage{color}
\usepackage{subcaption} 

\usepackage{amsmath,amssymb,amsfonts}
\usepackage{algorithmic}
\usepackage{graphicx}
\usepackage{textcomp}
\usepackage{booktabs}
\usepackage{cite}
\usepackage{url}
\usepackage{mathtools}
\usepackage{amsmath}
\begin{document}

\title{Networked Drones for Industrial Emergency Events }

\author{Maryam Khalid, Edward W. Knightly}

\maketitle

\begin{abstract}
	 Uncontrolled emissions of gases from industrial accidents and disasters result in huge loss of life and property. Such extreme events require a quick and reliable survey of the site for an effective rescue strategy planning. To achieve these goals, a network of unmanned aerial vehicles can be deployed that survey the affected region and identify safe and danger zones. Although single UAV-based systems for gas sensing applications are well-studied in literature, research on deployment of a UAV network for such applications, which is more robust and fault-tolerant,  is still in infancy. The objective of this project is to design the system that can be deployed in emergency situations to provide a quick survey and identification of safe and dangerous zones in a given region that contains a toxic plume without making any assumptions about plume location. We focus on an end-to-end solution and formulate a two-phase strategy that can not only guarantee detection/acquisition of plume but also its characterization with high spatial resolution. To guarantee coverage of the region with certain spatial resolution, we set up a vehicle routing problem. To overcome the limitations imposed by limited range of sensors and drone resources, we estimate the concentration map by using Gaussian kernel extrapolation. Finally, we evaluate the suggested framework in simulations. Our results suggest that this two phase strategy  not only gives better error performance but is also more efficient in terms of mission time. Moreover, the comparison between 2-phase random search and 2-phase uniform coverage suggest that the latter is better for single drone systems whereas for multiple drones the former gives reasonable performance at low computational cost.   
\end{abstract}


%
\IEEEpeerreviewmaketitle

\section{Introduction}

Industrial disasters such as fires, accidents and leakages etc. result in uncontrolled emission of toxic gases and chemicals in the atmosphere and affect large regions in their vicinity. The recent chemical plant fires in Houston are an example of such incidents. In such extreme events, sometimes the exact source of emission is known beforehand. However, mostly e.g in case of a leakage, the source location is not known. Regardless of whether the source information is available, in all such emergency situations, a reliable and quick survey of the area is required before starting the rescue operation. There have been several cases when rescue workers become victims during a rescue operation because of unidentified hazards and exposure to dangerous levels of poisonous gases\cite{emergency}. Thus, it is extremely important to collect information about toxic substances present in the affected area with high spatial and temporal resolution to identify hazardous regions.  \\
The deployment of sensing devices in extreme events not only saves human lives but also provides us high resolution information which would not have been accessible otherwise. This data can be leveraged to cope with the extreme situation more effectively and quickly. This strategy can also yield predictions about regions that might be at high-risk in near future and help in taking precautions to prevent further damage. Even if the event is not extreme, it is still important to identify regions where the air quality is being affected by the industrial emissions of pollutants, volatile organic compounds(VOC) and toxic gases.\\
This problem of measuring and estimating the distribution of toxic gases and VOCs can be classified under the umbrella of systematic environmental monitoring. There are three candidate technologies that have been explored in research and practice for gas sensing and environmental monitoring applications: wireless sensor networks (WSN), mobile robots and unmanned aerial vehicles (UAV) or drones. The deployment of wireless sensor networks (WSN) at fixed locations for gas leakage detection and air-quality monitoring is optimized offline before their deployment. Therefore, not
only the coverage is limited, it also doesn't cater to the dynamic nature of gas dispersion process that depends on numerous physical factors. These challenges are addressed by mobile robots but they are only suitable for indoor monitoring. In outdoor environments, these robots are constrained by limited mobility and uneven terrain. Moreover, their coverage is also limited to 2-D plane on ground. The most suitable candidate that is easy to deploy in emergency situations and guarantees coverage in 3-D and unconstrained mobility is unmanned aerial vehicle (UAV) or drone. The ability of drones to explore in 3-D provides better prediction accuracy. Furthermore, they easily access difficult-to-reach places like chimneys etc. \\
In this work we focus on network of cooperating drones instead of a single drone. It must be noted that compared to a single drone, a network is more flexible, robust, fault-tolerant and cost effective\cite{chemical_cloud}. The ability of a network to perform a task cooperatively and exhibit emergent behaviors makes it more flexible and robust than a single-UAV system. Moreover, if one drone fails the other can try to replace it by performing some part of its mission.\\
Along with drones, the sensing mechanism is also very important. Since, the gases are colorless, it is impossible to work with image-based sensors. Thus, we focus on gas sensors that measure the concentration of gas or VOC at the particular sampling location. It must be noted that these sensors need to come into contact with the gas/VOC in order to measure its concentration and therefore have essentially zero range around the sampling point. What further makes this problem challenging is the scale of affected area and drone's limited battery resource. The pollutants or gases emitted by a source are carried away to far away places by the wind and this dispersion can extend to order of kilometers around the source. This transport process is a complex process and it's hard to fit nice closed-form models to characterize this process. It depends on various physical factors such as temperature, humidity, wind, etc. and is inherently a random process because of turbulence and diffusion. Hence, instead of a model-estimation approach, a purely data-driven solution needs to be adopted for which real-time data with high spatial resolution is required. \\
As mentioned earlier, for extreme events that contaminate large parts of the environment with toxic substances in an instant, localizing the source of emission is not enough. Even though source location is a valuable, it is not enough to completely characterize the spatial extent of  contamination that has been done by that source. For such scenarios, the problem boils down to a coverage problem and identification of dangerous or highly contaminated regions within that covered area. We approach this problem with a 2-phase strategy.  We first \textit{explore} the area and conduct a coarse search for identification of high concentration region. In the second phase, the identified region is \textit{exploited} to classify safe and unsafe regions with high spatial resolution.\\
The rest of the paper is organized in the following manner. We describe the system components in section \ref{sys} followed by the related work in section \ref{prev}. In section \ref{sol}, we formulate the problem after which we discuss our strategy in section \ref{strat}. We evaluate the performance of our system through simulation in \ref{simulation} finally concluding in section \ref{conc}.

\section{System Description}\label{sys}
The system is compose dof three major blocks: toxic plume, gas sensing drone and spatial search space. There are certain specifications and assumptions associated with each block which we explain in following subsections.

\begin{figure}[t]
	\includegraphics[scale = 0.4]{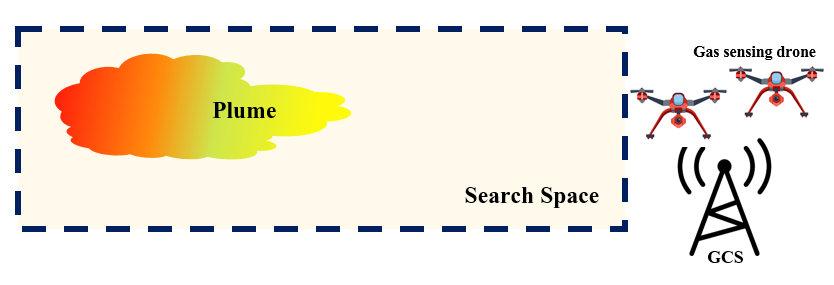}
	\caption{System Components}\label{system}
\end{figure}

\subsection{Toxic Plume}

A plume is a  mass of gas or pollutant coming out a source. The dark blackish-grey cloud coming out of chimney is one example of a plume. The plume has different zones based on the concentration of gas/pollutant and are usually labeled as low, moderate and high concentration regions. For ease of analysis and to avoid extra complexity, we refer to the high concentration (not safe) zone as the plume.\\
 Generally, when coming out from a point source, the gases/pollutants travel large distance along downwind direction while spreading in  in cross-wind directions and this spread becomes wider and wider as we move along the downwind direction. The transport process which is responsible for dispersion of these gases/pollutants, is a complex process that depends on a lot of physical factors and is therefore hard to model. Also, as it can be seen in figure \ref{system}, the plume has a highly irregular shape.  Therefore, fine resolution data is required for complete characterization of this plume. \\
For our problem, we define the plume mathematically as the concentration of gas/pollutant which is higher than a certain pre-determined threshold $C_{d}$. Usually this threshold is unsafe level of a particular gas/pollutant and therefore, a plume in effect characterizes the regions that are not safe.\\ 
Currently we are assuming that a single source continuous plume is present in the system. However, since we are focusing on overall gas distribution map instead of only source localization, our solution can be easily extended to multiple source system where the sources are independent of each other and are additive in their effect.

\subsection{Gas Sensing Drone Network}
The most important part of our system is the network of gas-sensing drones that performs the mission to achieve the required objectives. All drones fly from the ground control station (GCS) at the same time. However, during the mission, they operate in a completely autonomous manner and do not require control commands from the GCS. The data may be exchanged with the GCS to avoid high-computation and sophisticated estimation algorithms from running on the drones. The network performs the mission cooperatively and all drones can communicate to each other and the GCS.\\
 Moreover, each drone is equipped with a gas sensor. The gas sensor measures the concentration of gas/pollutant at that particular sampling location. Thus, the drone samples gas concentration values at all sampling locations along its path. The drones have limited flight time and fixed speed that specify the constraints of our system.   
 
\subsection{Spatial Search Space}
The third block of our system is the region defined as spatial search space. As mentioned earlier, wind acts as a carrier for transport of toxic substances. Therefore, generally speaking the scale of this space is of the orders of $m^2$ and can go up to $km^2$ which is huge given that the average flight time and speed of commercially available drones is 30 minutes and 20 $m/s$ respectively. \\
It is assumed the plume is fully or partially present in the defined space. The strategy laid out in this report still works even if plume is not present in the defined space. Once the exploration phase is complete and it is clear that plume is not present in the region, a new space can be defined and the process can be repeated. However, our strategy is not optimized for such scenarios without a preceding framework for optimal division of regions and allocation of drones. Thus, currently we focus on region that contains the plume and all the drones are allocated to it.\\
Although we assume that the plume is present in the pre-defined spatial region, we do not make any assumptions about the \textit{exact} location of plume inside this region. This feature is very critical to bring our system close to real-life scenario. Even though source locations are sometimes known, the continuous dangerous concentration zone inside the big plume is still unknown.\\
The search space and its characteristics play a significant role in providing guarantees and performance bounds specifying the reliability of  proposed setup.        

\section {Related Work}\label{prev}

The related work on UAV-based gas sensing and environmental monitoring can be divided into different categories. First of all, a major classification can be made on the basis of whether a single drone performs the job or a network of cooperating drones distribute the mission within themselves. In \cite{single_drone} and \cite{microdrone} a single drone is developed for gas source localization. The work in \cite{chemical_cloud} focuses on deployment of a network of drones for detection and mapping of a chemical cloud. In \cite{coverage} also, multi-UAV network is analyzed from the aspect of spatial coverage.
The second classification revolves around plume characterization. The initialization of drones from inside the plume or availability of prior information about the plume location can significantly reduce the required flight time. However, for most events like gas leakages the exact location of the source is not known. Even when the source information is available, the dangerous concentration regions inside the plume are still not known. Therefore, it is worthwhile to consider this plume acquisition phase where no prior information is available and the drones search for the plume. The work in \cite{single_drone} and \cite{coverage} assumes that the drones start from inside the plume. In \cite{microdrone}, the drone first searches for the plume, however the strategy for this phase is to follow a pre-defined path. Moreover, it is not specified \textit{how} that path is chosen or if it is optimal. \\
In \cite{chemical_cloud}, a random-search method is adopted to find the plume. This method can not provide any guarantees on the time required to find the plume or if it would be found within the flight time of drones. Also it suffers from high variance and is not suitable for emergency situations unless the search space is small and or the drone network is really large. Contrary to this, we adopt a systematic approach that not only guarantees acquisition of plume but also minimizes the mission time. \\
 After plume acquisition when the  drones start sampling non-zero or noticeable values of concentrations and are present near or inside the plume, the plume needs to be traversed in such a way that help achieve the desired goals. Although a significant amount of literature has focused on source localization, the idea of binary classification of regions into safe and danger zones has not received much attention. Even though source information can somewhat aid in achieving the latter objective, it can not completely characterize the extent of contamination mainly because of complex transport process and lack of models to perfectly characterize it.

\section{Problem Formulation }\label{sol}
The goal of this project is identification of region present in the search space that contains high level of toxic gases or VOCs. Since we are considering a single source continuous plume, it turns out that it is the plume that essentially defines the region we want to identify and mark as unsafe. Therefore, to formulate the problem our first step is to define how to characterize this plume-contained region. Based on this characterization, we compare the true distribution map of the area to the estimated one. The quality of our estimate or final classification depends on the information collected from the spatial region which is determined by the path taken by the drones. Thus, in the second step we focus on  data sampling and drone trajectory.

\subsection{Region Characterization}
\begin{figure}
	\centering
	\includegraphics[scale=0.35]{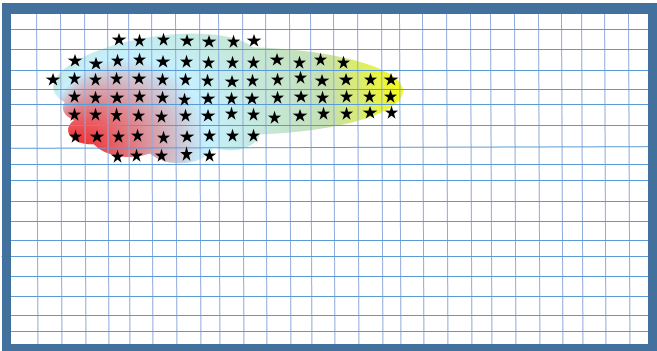}
	\caption{Distribution of region. The boxes containing black star mark unsafe region.}\label{smallb}
\end{figure}

To evaluate the performance of our system and characterize the irregular shape of plume, we divide the pre-determined spatial region into small 1x1$m $ squares as shown in figure \ref{smallb}. After the division, the maximum concentration value in each box is compared against a known safety threshold. If the value is above the threshold, the box is labeled as unsafe as can be seen by the starred boxes in figure \ref{smallb}. \\
 The maximum value can be replaced by the average value as well 
and this would reduce the overall number of boxes marked as unsafe specially the ones present at the plume boundary and are partially covered by the it. However,for the emergency situations that contaminate the area with toxic substances and it is more critical to mark a region as unsafe even if it partially contaminated. Thus, we choose  maximum value as the metric to compare against the threshold so that if the box is marked safe, it can be \textit{guaranteed} to be completely safe.  

\begin{figure}[b!] 
	\centering
	\includegraphics[scale=0.45]{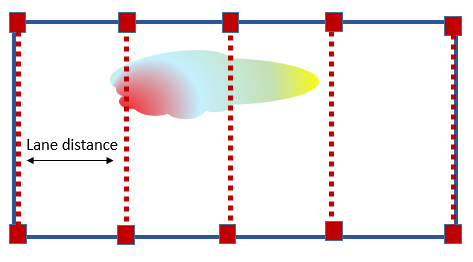}
	\caption{Spatial resolution for data collection}\label{lane}
\end{figure}

\begin{figure*}[t!]
	\centering
	\begin{subfigure}{0.4\textwidth} 
		\includegraphics[width=\textwidth]{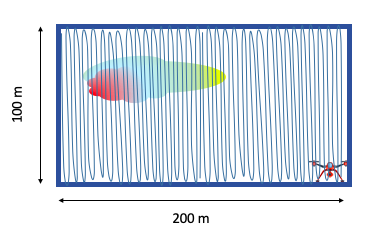}
		\caption{A fine resolution search scenario with lane distance = 2m.}\label{fine}
	\end{subfigure}
	\vspace{1em} 
	\begin{subfigure}{0.4\textwidth} \label{ideal}
		\includegraphics[width=\textwidth]{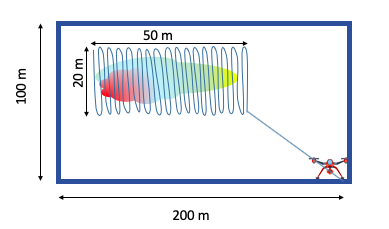}
		\caption{The ideal path scenario}\label{ideal}
	\end{subfigure}
	\caption{Two possible drone trajectories that yield zero estimation error}  
\end{figure*}
\subsection{Data Sampling and Drone Trajectory}
The solution adopted in this work relies strongly on the data collected by the drones during their trajectory. Given that the exact position of plume is not known, the complete region needs to be explored. To determine how finely the area is explored, we define the \textit{spatial resolution} with which the area is covered.
The area is transformed into a graph of nodes as shown by red squares in figure \ref{lane}. The lanes between corresponding boxes define the spatial resolution with which the area is traversed. Hence, the spatial resolution is inversely proportional to the lane distance.

\subsection{Problem Formulation}
In the previous subsections, we specified the tools needed for problem formulation. Now we formally state the objectives and define the problem.
Let $C_i^m$  represent the  maximum concentration present in box $i$ and $C_d$ represent the dangerous concentration threshold. The label assigned to the box is represented by $y_i$,

$$
y_i = I(C_i^m)
$$
where $I$ is an indicator function,
\begin{equation}
I(C_i^m)=\begin{cases}
1, & \text{if $C_i^m>=C_d$}.\\
0, & \text{otherwise}.
\end{cases}
\end{equation}

The performance is measured by the total misclassification error which is sum of false negative $FN$ and false positive error $FP$. To evaluate that, we represent the true and estimated labels by $y_i$ and $\hat{y_i}$ and the sets $\Omega_P$ and $\Omega_N$ represent the true distribution of boxes that are labeled $1$ and $0$ respectively. We can now compute the error as follows,

\begin{equation}
FN = \frac{1}{|\Omega_P|} \sum_{j\in \Omega_P} |y_j - \hat{y_j}|*100
\end{equation}

\begin{equation}
FP = \frac{1}{|\Omega_N|} \sum_{k\in \Omega_N} |y_k - \hat{y_k}|*100
\end{equation}

\begin{equation}\label{err}
\% Error = FN+FP
\end{equation}

Our objective is to minimize the total error specified in \eqref{err} by collecting data from multiple drones. This objective can be further broken down into smaller goals,
\begin{itemize}
	\item  First of all, it is required that the drones traverse the area and are able to detect the plume present in it. This can be called plume acquisition or detection and plays a crucial role in specifying the reliability of system for emergency situations .
	
	\item In the next step this plume needs to be characterized with high spatial resolution in order to estimate the gas distribution map of the area. In the final phase this map is used to identify unsafe zones.
	
	\item Since the focus of this work is on disaster response and emergency situations, it critical to achieve these objectives in minimum possible time. 
\end{itemize}
It can be observed that the error performance is directly dependent on the amount of non-redundant information that is collected by the drones which further depends on the drone flight time. Thus, the more time the drones spent int the spatial region in collecting diverse samples, the lower the error is. Thus, assuming that there are no revisits, there is a tradeoff between flight time and error. However, the urgent nature of scenario and the limited autonomy puts a constraint on the time that can be spent in collecting the data. Thus, our objective finally boils down to choosing that drone trajectory from the set of all the possible error-minimizing paths which takes the shortest time . 

\section{Proposed Strategy}\label{strat}
In this section, we explain our strategy and the intuition behind it to minimize the error. First of all consider a simple scenario in which the drones traverse the area with extremely fine resolution such that all the estimation error is zero.  Consider the scenario presented in figure \ref{fine} where a single drone covers a $200$ by $100m$ area with a resolution of $2m$ at a speed of $10 m/s$. In order to cover this region with such high resolution the drone needs to fly for approximately 17 minutes. \\

Consider the scenario presented in figure \ref{fine} where a single drone covers a $200x100m$ area with a resolution of $2m$ at a speed of $10 m/s$. In order to cover this region with such high resolution the drone needs to fly for at least 17 minutes without considering the sampling time. Now consider the other scenario presented in \ref{ideal}, in which the drone flies directly to the plume and flies over it to characterize it. If it traverses the plume with a resolution of 2 meter then the flight time is around 3-4 minutes. The estimation error in both scenarios is almost zero however the mission time is scenario 1 is four times more than that of scenario 2. Around three quarters of the total mission time in scenario 1 was spent in collecting redundant information and sampling low concentration locations that were not useful. The strategy adopted in scenario 2 is optimal in minimizing both, the estimation error and flight time. However it requires the exact location of plume and perfect sensors which makes it hard to deploy in practice. \\
Based on the features of both strategies mentioned above, it is intuitive to come up with a hybrid strategy that can leverage the benefits of both these strategies and provide reasonable performance in small time. Therefore, we consider a two phase strategy to balance the tradeoff between exploration and exploitation of the given region.
\begin{itemize}
	\item \textbf{Phase 1-Coarse Search and Estimation :} In the first phase, the drones traverse the area with low spatial resolution and the objective is to acquire the plume and have a coarse estimate of it's location and spread.
	
	\item \textbf{Phase 2-Fine resolution plume characterization :} In the second phase, the drones are directed to the region where the plume is estimated to be present. The data is collected with high spatial resolution in that region. Finally, the estimated concentrations are used to classify regions into safe and danger zones.
	
\end{itemize}
\begin{figure}
	\centering
	\includegraphics[scale=0.5]{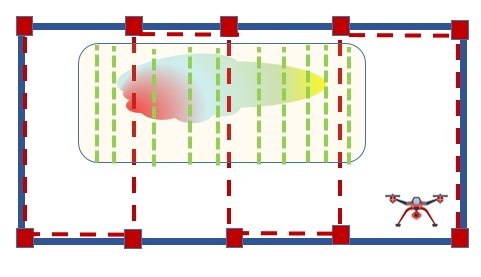}
	\caption{Two phase strategy for plume acquisition and characterization}\label{2phase}
\end{figure}
 An example of this strategy is shown in figure \ref{2phase}. The red lines represent the trajectory taken by drone in first phase. After the first phase, an estimate of high concentration region is computed to zoom into that region. The methodology adopted in both phases is explained in the following subsection.
 
 \subsection{Coarse Search and Plume Acquisition} The main objective of this phase is to not only detect or find the plume but also have a coarse estimate of its spread. As mentioned in section	\ref{prev}, the strategy adopted in \cite{chemical_cloud} was to set off all the drones in random directions and let them fly until any of them finds the plume and communicates that location to all others. This strategy is feasible if the proportion of plume region to the empty or low-concentration region is very large. However, if the search space is large, like the scenario we are considering, it is not feasible to use this random search method. Furthermore, this strategy can not provide guarantee that plume would be detected before the drones run out of battery. Since, a reliable and quick response is needed in emergency situations, it is not feasible to adopt random search strategy. \\
 \begin{figure}
 	\centering
 	\includegraphics[scale=0.4]{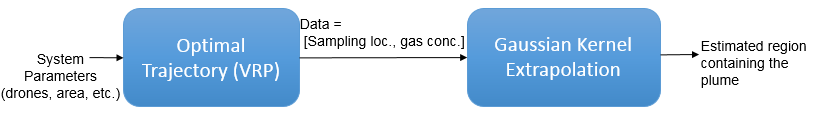}
 	\caption{Plume Acquisition block diagram}\label{acq}
 \end{figure}
 
 The other method is to define nodes and lanes as shown in figure \ref{lane} and ensure that drones ravel along these lanes. Since there are multiple ways in which multiple drones can cover the region with a given spatial resolution, the shortest path needs to be determined. For this we solve the vehicle routing problem (VRP)\cite{VRP} that finds the shortest path for a network of drones to visit a set of nodes while staying within the constraints imposed by the drone resources. After the drones have followed the trajectory computed by solving VRP and sampled data along its trajectory, the data is used to estimate the region where the plume is present. Therefore, the second block is estimation block for which we use Gaussian kernel extrapolation as shown in figure \ref{acq}.
 
 \subsubsection{Vehicle Routing Problem}
  
 To find the optimal trajectory, first of all the area is decomposed into set of nodes and lanes between parallel nodes. If the binary variable $x_{ij}^k$ represent whether the drone $k$ has traveled from node $i$ to $j$ and $v$ represents the drone speed, the flight time of drone $k$ is given by,
 
 \begin{equation}
 t_k = \sum_i \sum_j \frac{d(i,j)}{v} x_{ij}^k
 \end{equation}  
  
 where $d(i,j)$ represent the euclidean distance between node $i$ and $j$. To minimize the overall mission time, we need to minimize the maximum flight time. Thus, with area decomposed into $M$ nodes, the optimization problem with $N$ drones, each having battery life of $T_b$ minutes,can be formulated as follows\cite{VRP},

\begin{equation*}
\begin{aligned}
& \underset{X}{\text{minimize}}
& & U \\
& \text{subject to}
& & t_k \leq U, \; k = 1, \ldots, N.\\
&&& t_k \leq T_b, \; k = 1, \ldots, N.\\
&&& \sum_k \sum_j  x_{ij}^k = 1,  \;  j = 1, \ldots, M.\\
&&& \sum_i x_{ip}^k - \sum_j x_{pj}^k = 0, p = 1,\dots, M,\;  k = 1, \ldots, N.\\
&&&\sum_k x^k_{i,i+1} + \sum_k x^k_{i+1,i} = 1, \; i=2,4,6...,N.
\end{aligned}
\end{equation*}

The first constraint ensures that maximum flight time is minimized and the second constraint ensures that it is below the maximum flight time of drones. The third constraint ensures that each node is visited by no more than one drone. The fourth constraint  is required to make sure that the drone that arrives at a node is the same that leaves that node. Finally the last constraint enforces the drone that visits one node to visit its corresponding paralle node as well and thus make sure that the drones travel along the lanes defined by the nodes in figure \ref{lane}.
This optimization problem is solved in NP-hard and for simulation purpose it was solved in MATLAB using Yalmip toolbox and Gorubi solver \cite{VRP}. Once the trajectory was generated, data was collected with high sampling frequency along that path and fed to the estimation block.

\subsubsection{Coarse Estimation}
\begin{figure}
	\centering
	\includegraphics[scale=0.35]{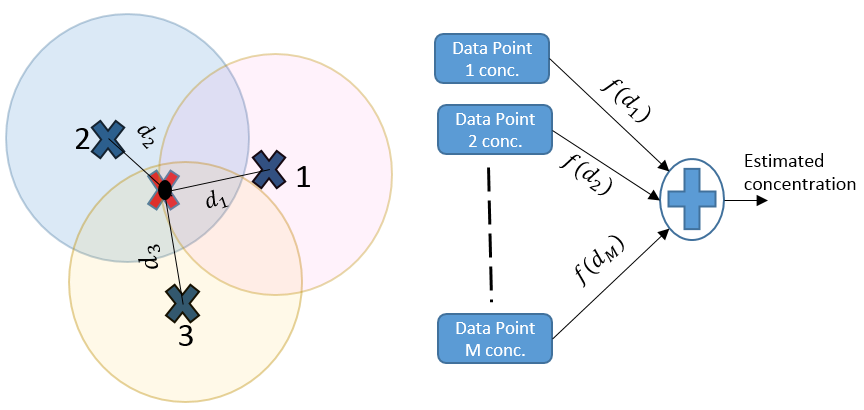}
	\caption{Gaussian Kernel Extrapolation}\label{gke}
\end{figure}

The gas sensors act as point sensors and therefore even with high resolution coverage, there are large regions for which no information is available. To make decisions about those regions, we need to leverage the sampled data. For this purpose, we use Gaussian kernel extrapolation. We have chosen this particular extrapolation method because of the following reasons, 
\begin{itemize}
	\item This method doesn't require even coverage of the area to make estimation. Hence, it can be deployed even with small amount of data.
	
	\item Since it integrates  multiple data points in a weighted average filter, it is robust to sensing noise and fluctuations.
	
	\item It is simple to compute and low computational complexity makes it efficient for deployment on the drone. 
\end{itemize}
Now we explain how this method works. First of all we define a radius $r$ for each sampled location to mark the spread in space where that particular point can have a contribution. Thus, for each sampled point we have a circle of radius $r$ as shown in figure \ref{gke}. Then, consider the position where the concentration needs to be determined. For that point, we consider all the data samples whose range/circle contain this point and represent them by set $\Omega_M$. In figure \ref{gke} the blue crosses show data points and  red cross represents the point for which we want to estimate. After determining the points in range, we take a weighted average of all those data values to estimate the concentration. If the estimation point is represented by $x$ and $C_k$ represents the concentration sampled at point $k$,  

\begin{equation}
\hat{C}(x) = \frac{1}{\sum_{k\in \Omega_M} f(d_k)} \sum_{k\in \Omega_M} f(d_k)*C_k
\end{equation}
where
\begin{equation*}
f(d_k) = \frac{1}{\sqrt{2\pi\sigma^2}} e^{\frac{d_k^2}{2\sigma^2}}
\end{equation*}

$$
d_k = ||x-k||_2
$$
The weights are gaussian functions of distance from the sampled point. The variance of this Gaussian function $\sigma^2$ is a design parameter and after choosing $\sigma^2$ the commonly used choice for radius is  $r=3\sigma$. It should be noted that choice of a large variance and therefore larger radius increases the tendency to estimate a larger plume that can result in higher false positive rate but reduce the risk of having false negatives. This parameter can be tuned according to the requirements of the system. \\
Once the concentration map for the complete region has been estimated, a square is wrapped around the estimated plume. Some additional margin is also added  to allow for estimation error. 
\subsection{Plume characterization and Region classification}

After the plume acquisition phase, a smaller region has been specified with a coarse estimate of the plume. The next step is to \textit{exploit} this region and get a better characterization of the plume. To achieve this goal, we look at two possible scenarios, random search and another coverage problem.

\subsubsection{Random search exploitation} In the second phase, we expect to start from inside or near the plume and therefore it is a good idea to leverage that information that is being sampled to guide the drones through the plume. Thus, the drones are initiated in random heading directions from the point with maximum estimated concentration. In order to ensure that multiple drones do not fly in the same direction unique intervals are defined from which the heading angle is selected randomly. The interval $I_k$ for drone $k$ is defined as follows,

$$
I_k = \Big[\frac{360(k-1)}{N},\frac{360k}{N}  \Big]
$$

where $N$ is the total number of drones.
Once the drones set off in random heading, each sampled concentration is compared against the threshold to determine if it has moved outside the plume. In case it moves out of the plume or the search space, a heading shift of $\pm$180\textdegree is added to bring it back into the region boundary or the plume. After a certain time interval the collected data from all the drones is combined with the data from phase 1 and is fed to Gaussian kernel extrapolation block to get the final estimate of concentration map of the hwole region.

\subsubsection{Uniform Coverage}
 Another approach is to repeat the procedure followed in phase 1 but with smaller area which is estimated to conatin the plume. Instead of initializing the drones from the ground station, they are initialized from the point with maximum estimated concentration. Another vehicle routing problem is solved to obtain drone trajectories.\\
After a certain pre-defined time interval, the collected data from all the drones is combined with the data from phase 1 and is fed to Gaussian kernel extrapolation block to get the final estimate for the concentration map of the whole region.

\begin{figure*}[t]
\centering
\includegraphics[scale=0.4]{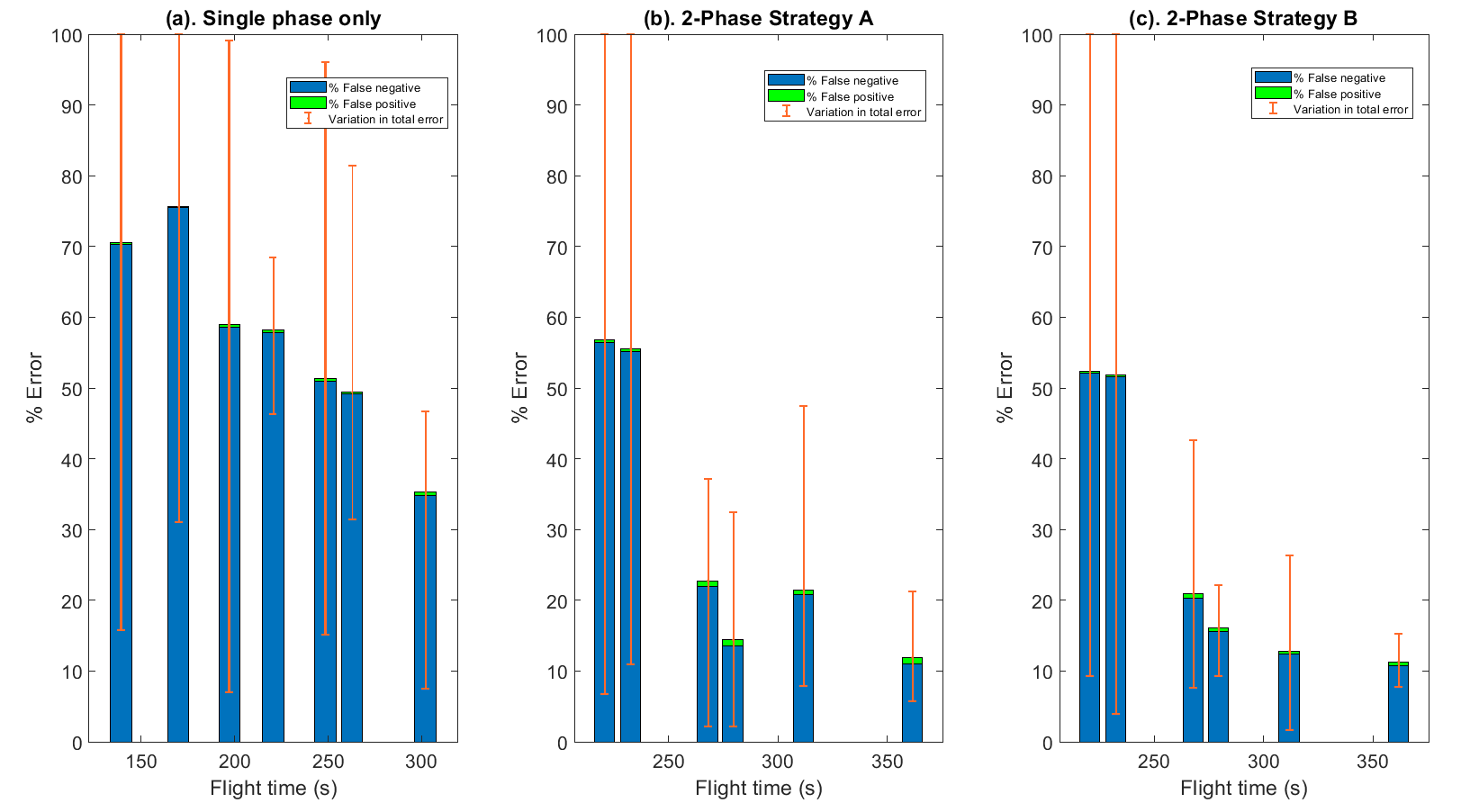}
\caption{Performance evaluation for single and two phase strategy with single drone}\label{res}
\end{figure*}
\section{Simulation and Results}\label{simulation}
We evaluate the performance of the framework discussed above through simulations. The objective of this simulation experiment is not only to see how well this framework minimizes the error but also to gain further insight about the following questions,
\begin{itemize}
	\item Does a two phase strategy promise improvement in average error and support the intuition behind it.
	
	\item If the error performance of first phase is good enough, is there still a significant gain in conducting the second phase.
	
	\item Since the plume location is unknown, it is expected that there would be high variability in the performance. How does this variance appear in both single and two phase strategy. 	
	 
\end{itemize}
\subsubsection{Methodology}
In order to answer these questions and have a performance evaluation, we design an experiment in the foll wing manner,
 
 \begin{itemize}
 	\item First of all we define the topology of our system. We consider a 200x100 m region with drones initializing from around the mid point of lower edge.
 	
 	\item After defining the topology, we generate synthetic data for gas plume by using Gaussian plume model. In order to incorporate the randomness introduced by unknown plume location, it is generated multiple times for different source locations and the results are averaged.
 	
 	\item The next step is to determine the fixed and changing variables. To observe the trends in error performance, the spatial resolution is varied over a range. The changes in spatial resolution directly reflect in flight time of the drones. Since time is a more crucial factor, we evaluate our performance against the time taken by the drones to complete the mission. Although it will result in non-uniform values on the x-axis, a big advantage of varying time by changing the spatial resolution is the assurance that time is not spent in exploring redundant information. Thus, an increase in time corresponds to \textit{more} area being explored. 
 	
 	\item After specifying all the design parameters, finally the VRP is solved to generate drone trajectory followed by data sampling and estimation. Error is computed for first phase followed by the second phase after which the final error is computed.
 	
 	\item Since multiple drones also add a gain in the performance, using it to compare the strategies might mask their true behavior and trend. Therefore, initially we perform the evaluation with single drone. After comparing the strategies, we look at the performance of multiple drones.
 \end{itemize}

\subsubsection{Results}
\begin{figure}
\centering
	\includegraphics[scale=0.5]{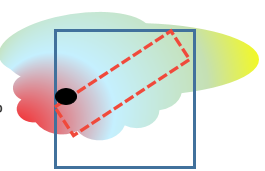}
	\caption{Random search revisit loop}\label{rev}
\end{figure}

The performance metric as defined in (\ref{err}) is composed of false positives and false negatives. As mentioned earlier, false negative is more critical than a false positive and therefore, we look at both of them separately along with the cumulative error.\\
The first plot on left of figure \ref{res} shows the variation in error as flight time increases for single phase strategy. The second plot is the result of two phase strategy with random search. The third plot is two phase with both phases solving the coverage (VRP) problem.\\
Starting from figure \ref{res}(a), it can be observed that in the beginning when flight time is small (less than 200 s) not only the error is very high but also the variance is very high and has a high tendency to miss the plume in acquisition phase as indicated by the errorbars going to 100.
As the flight time becomes greater than 200s, a downward trend starts to appear. Moreover false negatives constitute major part of the total error. The extremely small value of FP error can be attributed to the large proportion of safe region that is used to normalize it.\\
Moving to two phase strategy, in figure \ref{res}(b) and (c), the first two bars represent the scenario where there were plume instances that were not detected corresponding to $100\%$
error. The third bar and onward are more interesting. On average they do exhibit a downward trend however the random search strategy has some outliers.\\
Comparing the first plot with the other two, it is interesting to note that two phase strategy does provide a benefit. The performance achieved by phase 1 after 300 s is $35\%$ whereas the performance of two phase strategy after 270 s is around $20\%$. Thus, with two-phase strategy, there is not only a gain in performance but also in time efficiency. However, the gains are most significant when flight time for first phase is at the bottleneck where it detect the plume but the coarse estimate is really crude. In our case this point can be observed at 250 sec. With increasing flight times in phase 1, the gains from phase 2 start diminishing. \\
Now comparing the two phase strategies, on average B performs better than A because it guarantees better coverage of the estimated plume region. Moreover, the choice of random angle introduces high variance in performance for strategy A. \\
We also investigated the reason behind the outliers and high variance observed in second figure. It was observed that it had a tendency to stuck in loops and therefore suffered from problem of large revisits. An example is shown in  figure \ref{rev} where the initialized random angle and the structure of plume and defined area pushes the drone to get stuck in this loop. \\
Now we look at how much gain is provided in both strategies if multiple drones are used.
\begin{figure}
	\centering
	\includegraphics[scale=0.38]{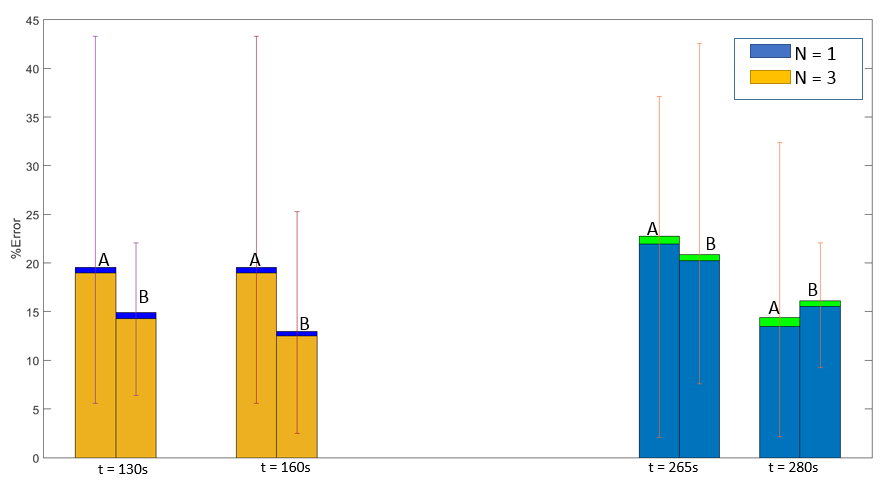}
	\caption{Comparison between single and multi-drone network}
	\label{result2}
\end{figure}
In figure \ref{result2}, the error performance for given flight time on x-axis is observed for a single drone and 3 drones. It can be observed that with single drone, although strategy B shows a uniform behavior, A does not. However with multiple drones, the behavior of A is much consistent. This is attributed to reduced variance in choice of random heading and diversity in the samples introduced by multiple drones exploring different locations. It can be concluded that with multiple drones not only there is a significant improvement in mission time but also the variations in random search strategy are reduced and it is much well-behaved and consistent in its performance.

\section{Conclusion}\label{conc}

In emergency situations, both quick and reliable survey of the affected area is required. UAVs network can provide these features only when the system design process focuses on both error minimization and time efficiency. The uncertainty introduced by unknown plume locations along with limited drone resources and limited sensing range of sensors further complicates the matter. To address these challenges, we focus on a two-phase setup with multiple strategies. The first phase can ensure plume acquisition and the second phase can help characterize it with high spatial resolution. The simulation results showed that there are significant gains achieved by deploying 2-phase strategy instead of just single phase. Although the 2-phase strategy with uniform coverage in second phase yields the best performance, solving the NP-hard coverage problem on the drones is not feasible. Compared to that, the random search strategy in second phase is much simpler. However, it is suffers from high variance and has a tendency to get stuck in revisit loops. Furthermore, it operates solely based on the data it collects and therefore is more prone to be affected by sensors' error. Thus, an effective strategy must be a hybrid approach that incorporates both data and coverage aspect to reach a balance between exploration and exploitation of the region. 

\bibliographystyle{IEEEtran}

\bibliography{my_bib}

\end{document}